\documentclass{article}
\usepackage{authblk} 
\usepackage{hyperref} 
\usepackage{amsmath}
\usepackage{graphicx}
\usepackage{float}

\def\nday{N_{day}}
\def\ncoun{N_{coun}}

\def\g#1#2#3{g_{#1}^{#2,#3}}
\def\s#1#2#3{s_{#1}^{#2,#3}}
\def\dd#1#2#3{dd_{#1#2,#3}}
\def\dab{D_{A,B}}
\def\wab{W_{A,B}}
\def\dmax{D_{\text{max}}}
\def\dmin{D_{\text{min}}}

\def\nday{N_\text{day}}
\def\ncoun{N_\text{coun}}
\def\ncont{N_\text{cont}}
\def\g#1#2#3{g_{#1}^{#2,#3}}
\def\s#1#2#3{s_{#1}^{#2,#3}}
\def\dd#1#2#3{dd_{#1#2,#3}}
\def\dab{D_{\!AB}}
\def\wab{W_{\!AB}}
\def\wabf{W_{\!AB}^{(5)}}
\def\dmax{D_{\text{max}}}
\def\dmin{D_{\text{min}}}
\def\stc{s_{_{\!T}}^c}
\def\beqa{\begin{eqnarray}}
\def\eeqa{\end{eqnarray}}
\def\a={&=&}
\def\ff#1#2{\includegraphics[scale=#1]{#2}}

\title{Global Streams, Local Currents:\\ A Data Analysis on Global VOD Content Consumption}

\author[1]{Nahyeon Lee}
\author[2]{Jongsoo Lim\thanks{\href{mailto:jslim123@sejong.ac.kr}{jslim123@sejong.ac.kr}}}
\author[3]{Mina Choi}
\author[4,5]{Hyeong-Chai Jeong\thanks{\href{mailto:hcj@sejong.ac.kr}{hcj@sejong.ac.kr}}}

\affil[1]{Department of Physics, Sejong University, Seoul 05006, Republic of Korea}
\affil[2]{Department of Media and Communication, Sejong University, Seoul 05006, Republic of Korea}
\affil[3]{Department of Media, Kyung Hee University, Seoul 02453, Republic of Korea}
\affil[4]{Department of Physics and Astronomy, Sejong University, Seoul 05006, Republic of Korea}
\affil[5]{Department of Computational Sciences, Korea Institute for Advanced Study, Seoul 02455, Republic of Korea}

\date{} 

\begin{document}

\maketitle

\begin{abstract}\label{sec1}

This study explores global video on demand content consumption patterns through a network-based approach.
We used Netflix's `TV-shows' ranking data, spanning 822 days across 71 countries, to construct a network where countries are represented as nodes and consumption similarities are reflected as link weights.
By applying the Louvain algorithm, we identified three distinct consumption groups, `North America and Pan-Europe', `Asia and Middle East', and `Central and South America group'. 
These groups align closely with geographic, historical, and linguistic divisions, despite no predefined grouping criteria.
Notably, Turkiye, often considered a cultural and regional crossroads, exhibited some classification ambiguity but was ultimately grouped with Asia and Middle East. 
Our findings also show that the United States accounts for the largest share of content consumption across all groups, while South Korean content, particularly after the success of ``Squid Game'' in 2021, has gained and maintained popularity in Asia, the Middle East, and Latin America.
This study, based on data, demonstrates that deep-seated cultural histories continue to shape global consumption patterns, even amidst rapid changes in media platforms and content production dynamics.

\end{abstract}

\section{Introduction}

Despite ongoing trends of de-globalization in politics and the economy, culture continues to globalize at an unprecedented pace.
Video on Demand (VOD) services, such as Netflix, have become major drivers of transnational popular culture.
While traditional global media consumption has often followed patterns shaped by cultural proximity\cite{straubhaar2007world}, viewing behaviors on VOD platforms may be influenced more strongly by geographic proximity.
Indeed, Netflix has managed its market by dividing it into four geographic cultural clusters, Europe-Middle East-Africa, Asia-Pacific, The United States-Canada, Latin America.

In spite of interest in VOD acceptance\cite{turner2019approaching}, empirical research became possible only after Netflix's TOP 10 consumption data was made public in 2020.
Lotz\cite{lotz2021between} has analyzed the frequency of overlap in the preferred Netflix Top 10 titles by major countries and argued that global consumption of Netflix consists of large, consistent markets such as English-speaking and Spanish-speaking regions, and small variable markets such as geographically close Korea, Japan, Taiwan, and the unique market of India. 
Jang et al.\cite{jang2023global} have identified eight groups that consume Netflix similarly based on Netflix's Top 10 consumption data, discovering that they are differentiated by geographical distance, linguistic similarity, and cultural differences.

The concepts of geo-cultural and geo-linguistic markets suggest that global VOD services are consumed differently across regions and languages.
Netflix’s 2021 diversity report emphasizes producing content that reflects regional cultures and inclusivity in race and gender\cite{netflix2021inclusion}, a strategy described as a deliberate effort to engage diverse audiences\cite{asmar2023streaming,lim2024is}.
Yet, important questions remain.
These include inquiries into how global VOD consumption is defined and separated, which nations are pivotal in this segmentation, and beyond it.
It is also essential to persistently monitor the evolving patterns of consumption among similar viewer groups over time with longitudinal data.
What distinct viewer groups emerge from the complex interactions of people connected to VOD platforms worldwide?

This study aims to examine how global VOD is segmented into national and regional markets.
These divisions are shaped not only by country borders but also by broader cultural and linguistic areas that often transcend national boundaries.
Essentially, we examine how the consumption of VOD services is influenced by and caters to these diverse geo-cultural segments.
To give a preliminary report, we found that global VOD content is being consumed in three distinct geo-linguistic consumer groups worldwide.
Some countries are located on the boundary of these consumption groups, with Turkiye being a prime example.
Due to its unique historical and geographical location straddling the East and West, Turkiye exemplifies a nation that could fall into more than one of these consumption
categories.

This may suggest the possibility that global audiences’ choices in VOD consumption are influenced not only by contemporary trends but also by historical and cultural dynamics.
The trajectory of VOD media consumption may be seen as reflecting a “long future”, a term used here to highlight the enduring influence of historical developments on present-day cultural practices.
Although not formally defined in this context, the phrase echoes Raymond Williams’s idea of a “long revolution”\cite{williams1961long} in cultural evolution, which he distinguished from more abrupt transformations like the Industrial or Civil Revolutions.
Williams argued that the cultural revolution unfolded gradually over centuries, culminating in the rise of mass media during the 20th century, long after the expansion of literacy and public readership.

In this light, today’s global VOD consumption patterns may represent a continuation of that cultural trajectory, where historical legacies still influence emerging forms of media engagement. This notion of a long cultural momentum underlines the interplay between global media flows and persistent local currents.

\section{Materials and methods}\label{sec2}

This research analyzes the top 10 Netflix show lists for each country from July 2020 to September 2022, to identify global trends and popularity patterns of Netflix Show (NS) and to find similarities in Netflix Show Consumption (NSC) patterns across countries.
By examining the differences in content rankings between countries, the `NSC distance' between two countries is calculated, where a shorter distance signifies a closer link between the two countries. 
We constructed a content consumption network with countries as nodes and classified the countries into groups based on their content consumption patterns using the Louvain algorithm\cite{blondel2008fast}.
The Louvain algorithm detects communities within a network by identifying groups with high intra-group edge weights and low inter-group edge weights, determining the optimal number of groups automatically instead of requiring user-specified group numbers.
We then determine the countries of origin for popular content within each group, identify the five nations most prolific in producing popular content for each respective group, and further quantify the degree of influence these leading five countries exert on the groups.
Finally, by measuring the popularity of content from these countries on a quarterly basis throughout the period, we examine the temporal trends and variations in the consumption share attributed to the top content-producing countries. 
Below, we provide a detailed account of the data utilized and the methodologies applied in conducting this analysis.

\subsection{Data}\label{subsec21}

We analyzed content lists of NS rankings, spanning from the 1st to the 10th place, collected from Flixpatrol over a period of $\nday=822$ days\footnote{July 1, 2020 to September 30, 2022} and across $\ncoun=71$ countries\footnote{The 71 countries mentioned are the sum of the members of the three groups listed in footnote 4.}.
If each content appeared in the list only once across the period, the total number of content entries would be $822\times71\times10=583,620$.
However, since most content appears in the list across multiple days and in different countries, the actual number of unique content entries was reduced to $\ncont = 1769$.

We denote the ranking of content $c$ on day $d$ in country $A$ as $\g{A}{d}{c}$.
Then country $A$ has a ranking dataset, 
\begin{equation}
    G_A = \{\g{A}{1}{1},\g{A}{1}{2},\ldots,\g{A}{2}{1},\ldots,\g{A}{\nday}{\ncont}\}
\end{equation}
composed of $822\times1769=1,454,118$ data points.
From this ranking data $\g{A}{d}{c}$ we introduce score data $\s{A}{d}{c}$ as follows.
\begin{equation}
  \s{A}{d}{c} = 
  \begin{cases} 
  11 - \g{A}{d}{c}, & \text{for } \g{A}{d}{c} \leq 10 \\ 
   0, & \text{otherwise}
  \end{cases}
  \label{e-rank}
\end{equation}
As equation~(\ref{e-rank}) indicates, the 1st place ranking is converted to 10 points, the 10th place to 1 point, and rankings beyond the 10th place are assigned 0 points.
By converting ranking data into score data, we can now represent each country as a point in a high-dimensional space of $822\times1769=1,454,118$ dimensions.
The coordinates of a country in this space are given by its score data.
For example, the position of country $A$ in this hyperspace is represented by the set of scores,
\beqa
S_A \a= \{\s{A}{1}{1},\s{A}{1}{2},\ldots,\s{A}{2}{1},\ldots,\s{A}{822}{1769}\}.
\eeqa
Within this coordinate system, we define the distance between two countries as the Euclidean distance in hyperspace, as detailed below.

\subsection{NSC distance between countries}\label{subsec22}

We aim to define the ‘NSC distance $D_{AB}$' between two countries as a measure of the difference in their content consumption patterns throughout the entire analysis period. 
For computational convenience, we first define the daily distance $dd_{AB,d}$ between countries $A$ and $B$ on day $d$, which is based on differences in their ranked contents.

\begin{equation}
  \dd{A}{B}{d} = \sqrt{\sum_{c=1}^{\ncont} {\left(\s{A}{d}{c} - \s{B}{d}{c}\right)^2}}
  \label{e-dis}
\end{equation}
The overall NSC distance $D_{AB}$ is then obtained by summing these daily distances over all days and taking the square root.
\begin{equation}
 \dab = \sqrt{\sum_{d=1}^{\nday} \dd{A}{B}{d}^2} 
    = \sqrt{\sum_{d=1}^{\nday} \sum_{c=1}^{\ncont} \left(\s{A}{d}{c} - \s{B}{d}{c}\right)^2}
 \label{e-disAB}
\end{equation}
In the summation, the terms $\left(\s{A}{d}{c} - \s{B}{d}{c}\right)^2$ calculate the squared difference
in scores between countries $A$ and $B$ for each piece of content $c$ on day $d$.
This approach ensures that all differences are positive and gives more weight to larger discrepancies, regardless of which country had the higher score.
Previous research\cite{jang2023global} has focused on whether contents were in the top 10 list or not, without considering their specific rankings.
We argue that taking ranking scores into account is crucial because the disparity between the 1st and 10th ranks is often more significant than between listed and unlisted content.
This is particularly true given the power-law like distribution observed in the popularity of cultural contents\cite{aggrawal2018view,clauset2009power,neda2017science} which suggests that higher-ranked content has disproportionately greater popularity and influence.

Although we cannot directly analyze the viewer distribution by rank due to the lack of access to actual viewership data per show, we analyzed a related quantity — the total score $\stc$ for each of the $\ncont$ contents — defined as
\beqa
 \stc \a= \sum_{k=1}^{\ncoun} \sum_{d=1}^{\nday} \s{k}{d}{c},
 \label{e.stc}
\eeqa
which sums the ranking scores of a given content $c$ across all countries and all days in the dataset.
We then ranked the contents by their total scores and plotted the distribution, which is shown in Fig.~A1 in the Appendix.
Although the distribution does not follow power-law in the entire region, but it clearly shows 
that a few contents received extremely high scores, while most received very low scores.
The gap between the most popular and the least popular content is very large, by a factor of about $10^5$ (five orders of magnitude).
This heavy-tailed pattern supports our assumption that higher-ranked shows tend to attract many more viewers than lower-ranked ones.

In equation~(\ref{e-dis}), the summation index goes from 1 to $\ncont =1769$, but in practice, we only add up at most 20 terms that are not zero. For example, if content $c$ is not in the top 10 for country $A$, then the corresponding term $\s{A}{d}{c}$ is zero, and the same rule applies to country $B$.
This implies that the only terms you need to consider are those where the content $c$ is in the top 10 for country $A$ or $B$. All other terms will be zero and can be ignored in the sum.

\subsection{NSC Network and Grouping}\label{subsec23}

Research on analyzing digital consumption data, such as that from Netflix, through network-based approaches has been extensively studied in recent  years\cite{aiello2012link,liu2016stability,abbas2018popularity,lobato2019netflix}.
In our study, we construct a consumption network where the nodes represent countries and the links between them are weighted by the similarity in content consumption between those countries.
Given the distances $\dab$ between each pair of countries, $A$ and $B$, we define the weights
$\wab$ as follows,
\begin{equation}
  \wab = \frac{\dmax - \dab}{\dmax - \dmin}
 \label{e-wab} 
\end{equation}
where $\dmax$ and $\dmin$ are the maximum and minimum values of $D_{ij}$ for all pairs of countries $(i,j)$ with $i \neq j$. 
Through this formula, the link between the two countries that are furthest apart in the hyperspace is assigned a weight of 0, indicating the lowest similarity, while the link between the closest countries receives a weight of 1, indicating the highest similarity.

Some may argue that this weight of the link between two countries  may not accurately capture the closeness in actual NSC patterns between two countries due to the limitations set by equation~(\ref{e-disAB}).
Specifically, one concern arises because the calculation considers only content ranked in the top 10, excluding any data on ranks below the 10th, which could potentially distort the measurement of consumption differences.
However, we believe this limitation is negligible due to the power-law distribution observed in content popularity\cite{aggrawal2018view,clauset2009power,neda2017science}, which suggests that the most significant insights come from the top rankings.

Due to the lack of data for ranks below the 10th, we cannot directly compare the weights calculated using all ranks to those calculated with only the top 10 content.
Instead, we assess the robustness of our approach by calculating weights $\wabf$ using only the top 5 ranked contents and comparing this order with the original weights $\wab$.
If these two sets of weights maintain a linear relationship, it confirms that analyzing the top 5 contents is sufficient, thereby supporting the validity of focusing on the top 10 list.

In Appendix, we show a scatter plot between two variables $\wabf$ and $\wab$ in Fig.~A2(a).
One can clearly see the linear dependence of these two weights.

Another concern arises from the fact that the release dates of content can vary between countries, and in some cases, certain content may be available in one country but not in the other.
To verify the validity of our methodology, we recalculated distances using only content that appeared in the top 10 on the same date in both countries.
This approach also yielded a strong positive correlation with the distances calculated using our original method, as detailed in the Appendix.
Therefore, we are confident in the validity of our analysis results despite the limitations of approximating countries as points in a hyper-dimensional space and using Euclidean distance to represent differences in VOD consumption across countries.

With this network, we classify the countries into NSC groups using the Louvain algorithm\cite{blondel2008fast}, a method widely applied across various fields for community detection\cite{blondel2015survey,mattei2021italian,lee2022network,betancourt2023temporal,cho2023multiresolution,kalhor2024understanding}.
The Louvain algorithm detects communities by maximizing the modularity $Q$, which is defined as
\begin{equation}
    Q = \frac{1}{2m} \sum_{AB} \left[ \wab - \frac{k_A k_B}{2m} \right] \delta(c_A, c_B)
\end{equation}
Here, $m$ represents the sum of the weights of all links, $k_A$ and $k_B$ are the sums of the weights of the links connected to countries A and B respectively, $c_A$ and $c_B$ denote the groups to which countries A and B belong, and $\delta(c_A, c_B)$ takes a value of 1 if the two countries belong to the same group, and 0 if they belong to different groups.
Due to the stochastic nature of the Louvain algorithm in determining the sequence for evaluating nodes for possible reallocation to different groups, we conducted 1,000 simulations to mitigate this variability.
Based on these simulations, we assigned each country to the group to which it most commonly was affiliated, thus establishing groups of countries with similar VOD consumption patterns.

\section{Results}\label{sec3}

\subsection{VOD similar consumption groups}\label{subsec31}

\begin{figure}[!h]
\centering
\ff{.6}{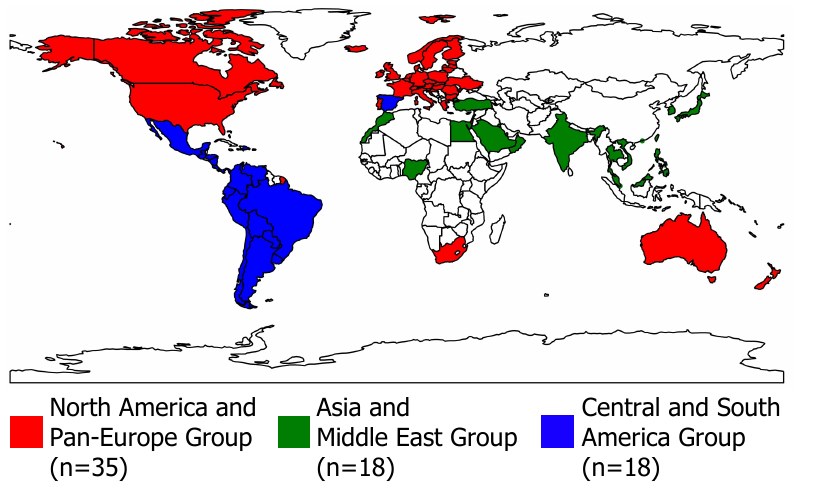}
\caption{VOD Similar Consumption Groups.
  The North America and Pan-Europe group (n = 35) is color-coded in red, the Asia and Middle East group (n = 18) is color-coded in green, and the Central and South America group (n = 18) is color-coded in blue.}
\label{fig:figure1}
\end{figure}

Figure~\ref{fig:figure1} shows the countries grouped by the Louvain algorithm from the data of 1,769 pieces of content, indicated by colors.
In all 1,000 simulations, the countries of 71 nations were classified into three groups, and the countries belonging to each group were always consistent except for one country. Looking at the countries in each group, it can be confirmed that they are classified into three VOD consumption groups\footnote{
North America and Pan-Europe Group: Australia, Austria, Belgium, Bulgaria, Canada, Croatia, Cyprus, Czech Republic, Denmark, Estonia, Finland, France, Germany, Greece, Hungary, Iceland, Ireland, Israel, Italy, Latvia, Lithuania, Netherlands, New Zealand, Norway, Poland, Portugal, Romania, Slovakia, Slovenia, South Africa, Sweden, Switzerland, Ukraine, United Kingdom, United States\\
Asia and Middle East Group: Egypt, Hong Kong, India, Japan, Kuwait, Malaysia, Morocco, Nigeria, Oman, Philippines, Qatar, Saudi Arabia, Singapore, South Korea, Taiwan, Thailand, Turkiye, Vietnam\\
Central and South America group: Argentina, Bolivia, Brazil, Chile, Colombia, Costa Rica, Dominican Republic, Ecuador, Guatemala, Honduras, Mexico, Nicaragua, Panama, Paraguay, Peru, Spain, Uruguay, Venezuela.
}:
the ‘North America and Pan-Europe Group,’
the ‘Asia and Middle East Group’, and
the ‘Central and South America Group’.

The only exception was Turkiye. Out of 1,000 instances, Turkiye was classified into the Asia and Middle East group 749 times, and into the North America and Pan-Europe group 251 times.
This reflects its geographical position at the border of Europe and Asia and its culturally integrative character. The fluctuation illustrates cultural hybridization, where overlapping ties lead to community inconsistency in Turkiye’s group assignment.
Turkish TV also exerts influence in a variety of markets from the Middle East to Europe and Central Asia\cite{kaptan2020television}.
Overall, global VOD consumption appears to be strongly associated with geographical, linguistic, and historical factors.

To address the robustness of group detection, we examined the distribution of the modularity score $Q$ across the 1,000 runs. When Turkiye was assigned to the Asia and Middle East group, the modularity was $Q = 0.12292$, and when assigned to the North America and Pan-Europe group, it was $Q = 0.12299$, showing no significant difference.
Although these $Q$ values are below 0.2, a threshold often cited as indicating weak community structure, this outcome arises from the all-to-all nature of our network, which tends to lower
absolute $Q$ values.
To assess the meaningfulness of the grouping, we generated random groupings with the same group sizes (35, 18, and 18 countries) and calculated $Q$ 1,000 times.
The random groupings yielded an average $Q = -0.009193$ with a standard deviation of $0.003706$, giving a $z$-score of $35.64$ for our observed result.
This large positive $z$ demonstrates that, despite the low absolute $Q$ values, the detected grouping is highly significant compared with random partitions.

We also note that our results are not sensitive to the specific definition of similarity weights between countries.
As shown in Appendix~A2, different definitions of similarity weights exhibit a strong linear relationship with each other, and the resulting group classifications remain largely consistent across these alternatives.
In particular, when cosine similarity is used, which depends only on the angle between country vectors in the hyperspace rather than their Euclidean distance, the analysis produces the same three VOD similarity groups with identical group memberships as those obtained by our original method.
This indicates that the observed group structure reflects stable patterns of VOD consumption and is not strongly affected by the choice of similarity metric.

\begin{figure}[h]
\centering
\ff{.51}{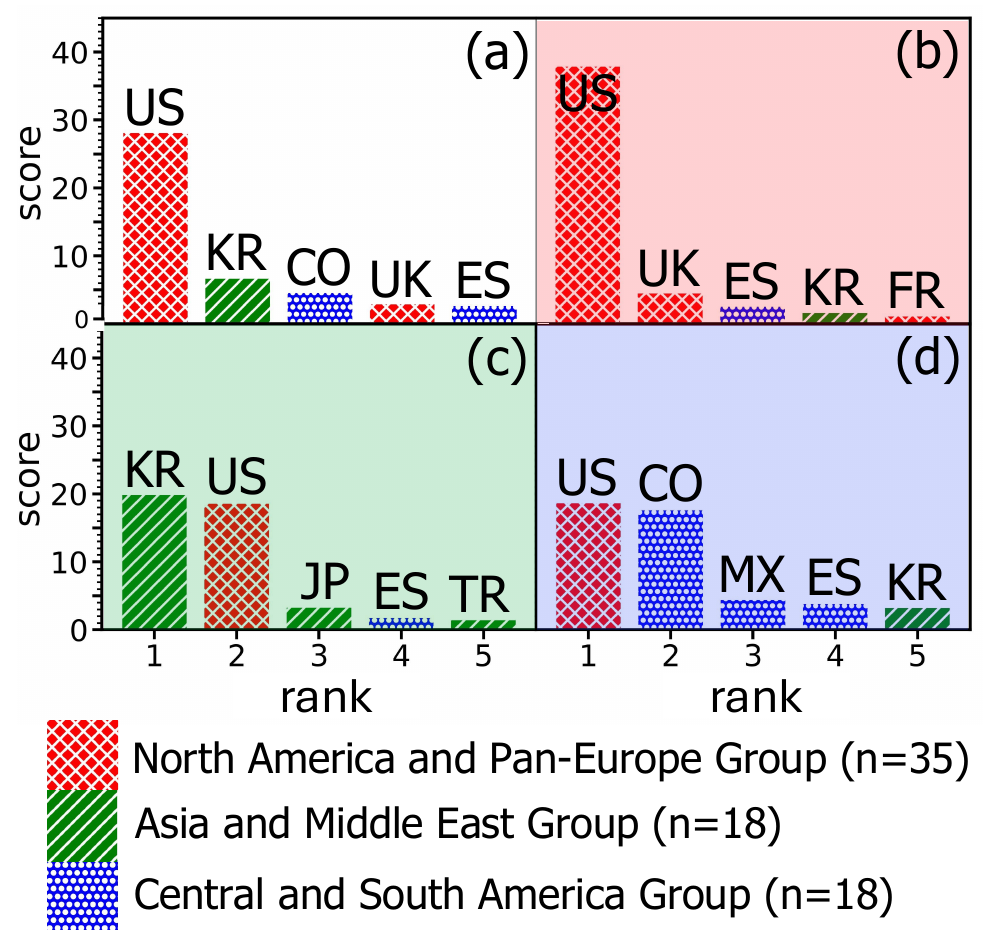}
\caption{
  Top 5 content producing countries and their scores within each group.
  The background color of the graph represents the group being averaged, and the color of the hatched bars represents the group that the producing country belongs to. ‘US’ stands for the United States, ‘KR’ for South Korea, ‘CO’ for Colombia, ‘UK’ for the United Kingdom, ‘ES’ for Spain, ‘FR’ for France, ‘JP’ for Japan, ‘TR’ for Turkiye, and ‘MX’ for Mexico.}
\label{fig:figure2}
\end{figure}

The dataset contains a total of 1,769 titles produced across 49 countries.
The production country of each title was identified using metadata provided by FlixPatrol, which classifies origin based on publicly available information such as the country of the main production company or the official release country.
To evaluate how much content from each producing country was consumed globally, we define a production score for each country.
The production score of country $p$ is defined as the sum of the total scores of all contents produced in that country,
\beqa
S_p \a= \sum_{c \in C_p} s_T^c ,
\eeqa
where $C_p$ is the set of contents produced in country $p$, and $s_T^c$ is the total score of content $c$ as defined in Eq.~(\ref{e.stc}).
For instance, the production score for the United States is obtained by summing the total scores of 528 US-produced titles, while that for South Korea is based on 203 South Korean titles.
This measure provides a way to compare the relative global presence of producing countries in terms of VOD consumption.

Since each country has contents ranking from 1st to 10th place daily, converting these to scores and summing them up gives 55 points.
That is, if all content from 1st to 10th place in a particular country on a particular date was produced by the United States, then the score for the United States in that country on that date
would be 55 points.
We calculated the average scores for 71 countries over 822 days.
The top 5 producing countries with the highest average scores are shown in
Fig. \ref{fig:figure2}-(a). Figure \ref{fig:figure2}-(b),(c) and (d) show the results of calculating the average scores for each of three VOD similar consumption groups.
Figure \ref{fig:figure2}-(a) represents the average scores over 822 days for all 71 countries, Fig. \ref{fig:figure2}-(b) for the 35 countries in the North America and Pan-Europe group, Fig. \ref{fig:figure2}-(c) for the 18 countries in the Asia and Middle East group, and Fig. \ref{fig:figure2}-(d) for the 18 countries in the Central and South America group.
The background color of the graph shows the group over which the average is taken, while the color of the hatched bars shows the group of the producing country.

The United States, which ranks first among all countries (see Fig. \ref{fig:figure2}-(a)), occupies high ranks of first, second, and first respectively in the three groups (each in Fig. \ref{fig:figure2}-(b), (c), (d)).
South Korea and Spain, which rank second and fifth overall, are also in the Top 5 across all three groups, indicating that content from the United States, South Korea, and Spain consistently receives global approval as a constant variable.
In contrast, Colombia and the United Kingdom, which rank third and fourth overall, are localized to the Central and South America group and the North America and Pan-Europe group respectively, unlike the aforementioned three countries.

Besides the constant variables of the United States, South Korea, and Spain, the countries that rank highly within each group include the United Kingdom and France in the North America and Pan-Europe group, Japan and Turkiye in the Asia and Middle East group, and Colombia and Mexico in the Central and South America group.
Interestingly, all of these are countries whose background color and bar color match, that is, they belong to their respective groups.
This shows that countries tend to consume a lot of content from countries within the same group.
Ultimately, the United States, South Korea, and Spain are characterized by content that attracts audiences across multiple groups, whereas the content from other countries is primarily consumed within their own group.

\subsection{The role of key content-producing countries in similar consumption grouping}\label{subsec32}

\begin{figure}[!htbp]
\centering
\ff{.52}{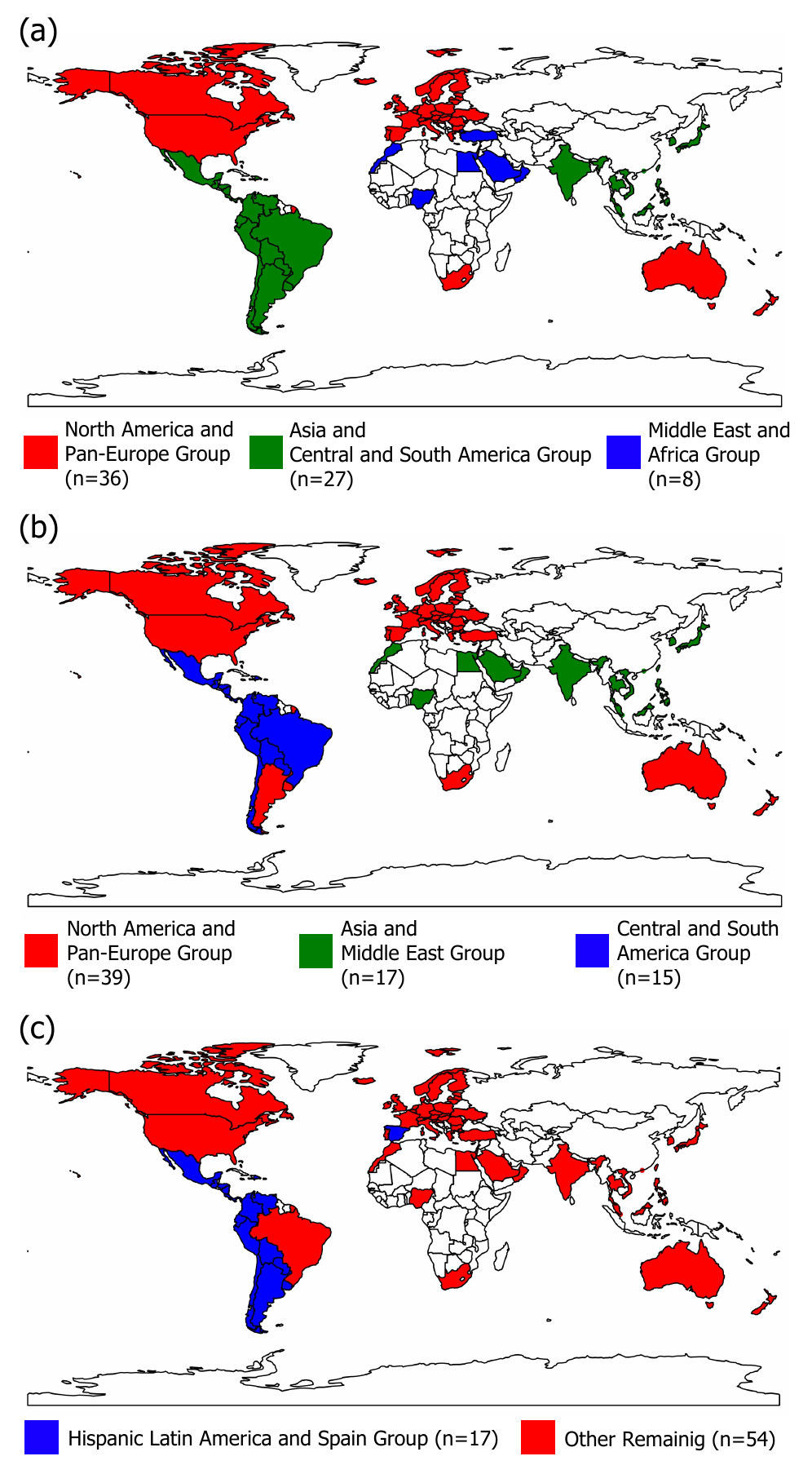}
\caption{Similar consumption groups by the leading content-producing countries.
  (a)Similar consumption groups by US-produced content,
  (b)Similar consumption groups by South Korea-produced content,
  (c)Similar consumption groups by Colombia-produced content
}
\label{fig:figure3}
\end{figure}

Then, to what extent do the key content-producing countries influence the grouping into similar consumption clusters?
To examine the extent to which major content-producing countries are associated with the formation of consumption clusters, we conducted the Louvain algorithm process again, but this time we only included the consumption data from the leading content-producing country within each of the three identified consumption groups.
Networks were constructed from 528 pieces of United States content, 203 pieces of South Korean content, and 27 pieces of Colombian content.

When forming similar consumption clusters based on the leading countries’ content, most countries consistently belonged to the same group across 1,000 runs of the Louvain algorithm, similar to the previous results. 
A few countries, however, were classified into different groups in some runs. 
Even so, every country was assigned to one group at least 700 times out of 1,000, and Fig.~\ref{fig:figure3} shows the groups to which countries most frequently belonged.

As seen in Fig. \ref{fig:figure3}, we can observe slight changes in the composition of the earlier three similar groups when we form similar consumption groups around specific country-produced content.
When focused on US content consumption, the North America and Pan-Europe cluster remained unchanged.
However, the countries in the Asian and Middle East group, as well as the Central and South America group, get reorganized.
The newly formed groups are the Asia and Central South American group and the Middle East and Africa group.
Within this new classification, Spain, which has previously been in the Central and South America group, is now grouped with the North America and Pan-Europe group. 
Turkiye, which has previously been in the Asian and Middle East group, is classified into the Middle East and Africa group. 

For South Korean content, representing Asia and the Middle East, the content consumption patterns and the overall grouping were similar to the original three groups that were previously identified.
Yet, notable differences were observed.
For example, Argentina, which has been grouped with the Central and South America group when considering a wide range of content, is now classified alongside North American and Pan-European countries when focus is narrowed to South Korean content.
Additionally, both Spain and Turkiye are categorized within the North American and Pan-Europe group. 

Finally, when the content consumption patterns are analyzed with Colombian content as the focus, we observed two main cluster groups, a Hispanic Latin America and Spain group and other
remaining groups.
That is, when looking from the perspective of Colombia, the division of groups appears to be heavily influenced by language, specifically whether the countries are Spanish-speaking or not.
This reflects the cultural and linguistic influence on media consumption within the VOD market. 

In sum, US content is similarly consumed in Asia, Central and South America, as well as the Middle East and Africa.
Specifically, Spain is associated with the North America and Pan-Europe consumption group when it comes to US content, while Turkiye is associated with the Middle East and Africa.
This indicates that audiences in Spain and Turkiye share consumption patterns of US content with North America and Pan-Europe, and with the Middle East and Africa, respectively.
South Korean content consumption shows patterns that closely resemble the three consumption clusters previously identified.
However, countries such as Argentina, Spain, and Turkiye display viewing patterns of South Korean content that are more similar to those of North America and Pan-Europe.
Colombian content is mostly consumed within the Spanish-speaking cultural sphere, including Latin America’s central western regions and Spain in Europe,
while other countries are grouped into a different cultural sphere.
Overall, this analysis suggests that content from major producing countries is associated with distinct global consumption patterns, which may converge or diverge depending on cultural and linguistic ties.
The way content from these countries is consumed may therefore contribute to shaping or reflecting cultural spheres and clusters within the VOD space.

\subsection{Trends in content-producing countries}\label{subsec33}

\begin{figure}[!htbp]
\centering
\ff{.52}{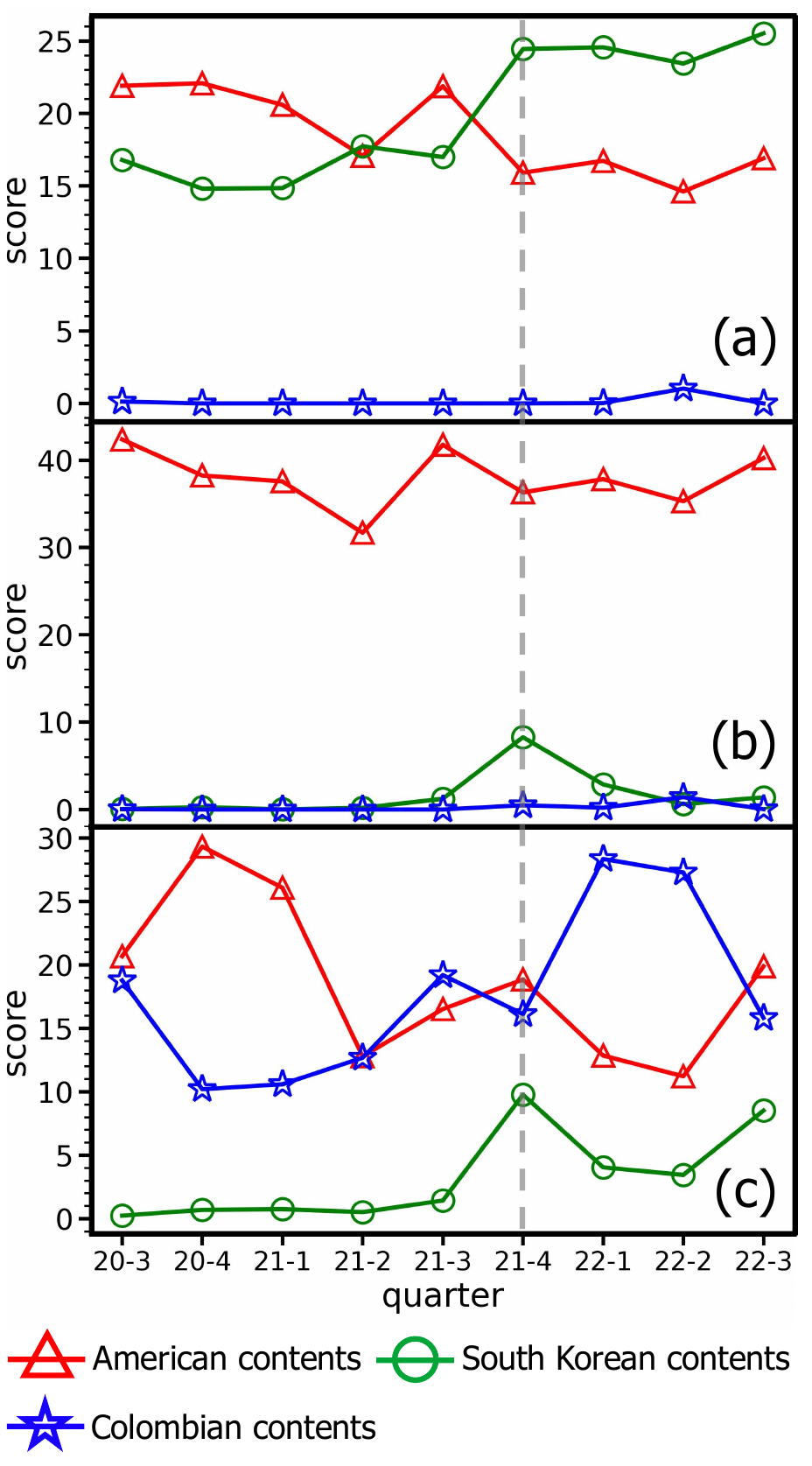}
\caption{
  Trends in the rankings of US, South Korean, and Colombian content by similar consumption groups.
  The red triangle represents the scores for American content, the green circle represents South Korean content, and the blue star represents Colombian content.
  (a)The Asia and Middle East Group,
  (b)The North America and Pan-Europe Group,
  (c)The Central and South America Group.
  }
\label{fig:figure4}
\end{figure}

Based on the original analysis, which defined three similar consumption groups by considering all content, we further examined how the content rankings of leading countries within the VOD similar consumption groups, such as the US, South Korea, and Colombia, have changed over the data period.
This aims to assess whether the previously observed content performance reflects random variation or consistent patterns.
Figure \ref{fig:figure4} represents graphs that illustrate the quarterly ranking scores of content from the United States, South Korea, and Colombia within the three established similar consumption groups.

A notable peak in the preference for South Korean content appears in the fourth quarter of 2021 and has continued to grow steadily thereafter. 
This pattern may be partially associated with the global popularity of Korean series released during that period, including Squid Game, which became a worldwide phenomenon.
In the North America and Pan-Europe group, while American content remains strongly preferred, there is a noticeable spike in interest for South Korean content in the fourth quarter of 2021, which coincides with the release period of Squid Game, followed by a return to prior levels.
In the Central and South America group, South Korean content consumption, which was previously minimal, shows a noticeable increase in the fourth quarter of 2021—a period that, once again, overlaps with the global release of Squid Game—and remains at a higher level thereafter.

The global response to South Korean content shows a noticeable change after 2021, during the same period as the release of Squid Game, which may have contributed to the shift.

\section{Summary and Concluding Remarks}\label{sec4}

Our findings show that global countries are categorized into three groups with geographic and linguistic connections based on global VOD consumption distance, the ‘North America and Pan-Europe group’, the ‘Asia and Middle East group’, and the ‘Central and South America group’.
In these three groups, the ‘North America and Pan-Europe group’ is most influenced by the United States, the ‘Asia and Middle East group’ by South Korea, and the ‘Central and South America group’ by the United States and Colombia.

Looking specifically at the top five producing countries, the United States shows the highest share of production scores in all three groups, followed by Spain and South Korea, which contribute at intermediate levels.
Other countries also stand out depending on their economic and cultural presence within the respective groups.
This pattern partially aligns with the fact that Netflix’s original investment in Europe is made selectively according to existing cultural power\cite{afilipoaie2021netflix}.

It highlights that the United States accounts for the largest share of content consumption, the distinctive geographical and linguistic reach of Spain, and the role of countries with notable production shares within each group.
South Korea is an exceptional case that cannot be explained solely by geography or language within this framework, warranting further examination.

The rise in South Korean content popularity observed in the fourth quarter of 2021 coincides with the release of the highly popular series Squid Game.
After this period, the preference for South Korean content consistently increased in the ‘Asia and Middle East group’ and the ‘Central and South America group’.
In contrast, in the ‘North America and Pan-Europe group’, the preference for South Korean content subsided relatively quickly, likely reflecting the strong presence of US content in that region.
These observations suggest that the release of widely recognized content such as Squid Game may be associated with shifts in global consumption patterns, even in regions with established viewing tendencies.
The increased attention to South Korean content following Squid Game indicates a notable change in its global visibility and consumer appeal, with especially strong gains in Asia and meaningful presence in the Central and South American markets.
While this surge in popularity appears to diminish more rapidly in regions such as North America and Pan-Europe, the broader trend points to an expanding role of  South Korean media within global VOD consumption, highlighting the fluid and dynamic nature of content preferences across cultural spheres.

It is noteworthy to mention Turkiye’s unique position in global media consumption.
Turkiye occupies the most ambiguous position among the three similar groups.
Data suggests that Turkiye consumes American content from the perspective of the Middle East and Africa group, whereas it consumes South Korean content from the perspective of the
North America and Pan-Europe group.
Such results suggest that Turkiye’s content preferences align variably with different cultural groups depending on the content’s origin.

Viewing users’ engagement with global VOD platforms within a broader anthropological and cultural trajectory, Raymond Williams explored the gradual progression of cultural revolution in his pivotal work ``Television: Technology and Cultural Form''
published during the height of the mass media era in 1974.
Today, about fifty years later, the widespread consumption of VOD content within this long historical and cultural trajectory offers new opportunities for researchers to explore.

The consumption of global VOD content seems to manifest what Williams might call a ‘long future’, a notion that emphasizes the gradual, cumulative processes of cultural transformation rather than a fixed or distant endpoint.
Yet, it remains to be observed whether the recent success of unique cases such as South Korea, with hits like Squid Game and Pachinko, is an accidental ‘outcome’ or will prove to be a sustained ‘factor’ in the future.

In conclusion, our study examines global VOD media consumption and identifies clear patterns shaped by cultural, geographical, and linguistic associations\cite{brown1991human}.
Using similarity networks, we find that countries consistently cluster into three major groups, reflecting both regional proximity and shared cultural characteristics.
The analysis also shows that content from major producing countries, such as the United States, South Korea, and Colombia, is associated with distinctive global consumption patterns, which may converge or diverge depending on historical and linguistic ties\cite{sommier2014concept}.
These results highlight how VOD platforms, while transcending national boundaries, still reflect underlying cultural structures and offer a framework for understanding the ways in which global audiences organize around shared viewing preferences.

Future research should examine whether current shifts, such as the growing popularity of non-Western content, will become lasting features of the global media landscape or prove to be short-lived.
In addition, it would be an intriguing topic to explore not only the contemporaneous relationships among countries but also the temporal dynamics of content spread across nations\cite{lee2025patterns}.
VOD media, positioned at the intersection of technology, culture, and audience choice, will remain a key area for studying how global and local cultural forces interact over time.

\appendix
\renewcommand{\thesection}{Appendix~\Alph{section}} 
\renewcommand{\thesubsection}{\Alph{section}.\arabic{subsection}} 
\renewcommand{\thefigure}{\Alph{section}\arabic{figure}} 
\setcounter{section}{0}
\setcounter{figure}{0}

\section{Validation of the Method}

In this appendix, we examine the validity of our methodology and test whether our results are robust under different assumptions. 
As explained in the main text, our analysis is based on Netflix Top-10 show rankings for each country from July 2020 to September 2022.  
The dataset contains
$\ncont=1769$ unique contents recorded across
$\nday=822$ days.  
Rankings were converted into scores, with 10 points assigned to 1st place, 9 points to 2nd place, and so on down to 1 point for 10th place, while contents outside the top-10 list received 0 points.  
Each country was then represented as a point in a high-dimensional space of $822 \times 1769 = 1,454,118$ dimensions, with coordinates determined by these scores.  
Distances between countries were calculated using the Euclidean distance, and then converted into weights $\wab$ using Eq.~(\ref{e-wab}) in the main text.  

This method provides a systematic and interpretable way to measure similarities in consumption patterns.  
However, two potential limitations require further validation.  
First, the availability of specific content can differ across countries.  
Second, our analysis uses only the top-10 rankings, which means we cannot distinguish between contents that were unranked and those that narrowly missed the top-10.  

To address these issues, we test whether restricting the analysis to the top-10 is a valid approach and then explore alternative definitions of distance between countries.  
By doing so, we assess whether the similarity patterns observed in the main text remain robust under these variations.

\subsection{Distribution of Content Scores}

\begin{figure}[t]
\centering
\ff{.46}{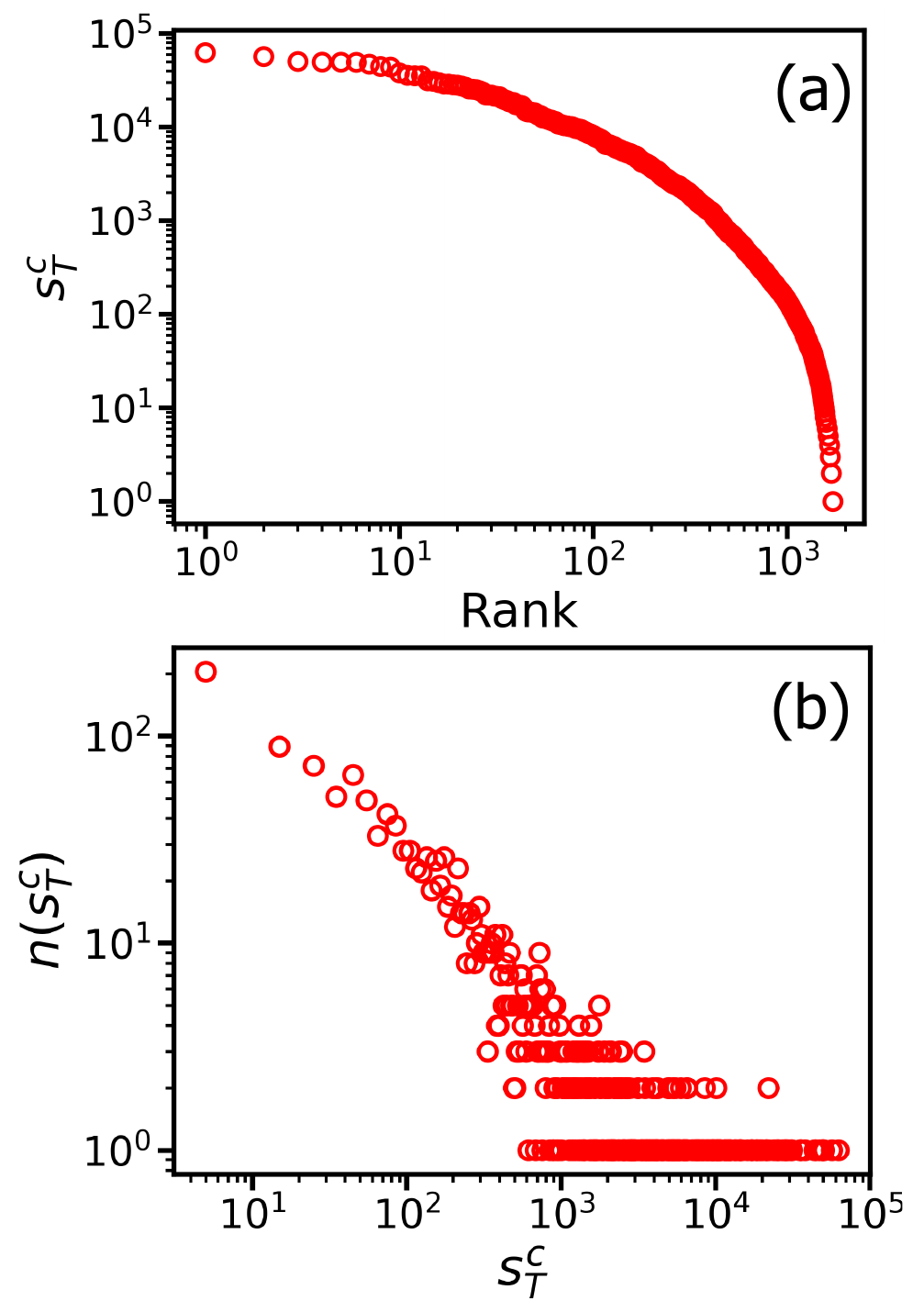}
\caption{
  (a) Total scores $\stc$ of $\ncont = 1769$ Netflix titles arranged by rank,
  plotted on a log--log scale.
  (b) Distribution of total scores $\stc$ on a log--log scale using bins of width 10.
  The vertical axis indicates the number of contents $n(\stc)$ falling within each bin.
}
\label{fig:figureA1}
\end{figure}

In this section, we examine the distribution of content scores to understand the variation in popularity among Netflix shows.  Since direct viewership data are not available, we use the total score $\stc$ defined in Eq.~\eqref{e.stc} as a proxy for global popularity.  
This quantity sums the ranking scores of each content across all countries and days in the dataset.

Figure~\ref{fig:figureA1} presents two complementary views of this distribution.  
Panel (a) shows the total scores $\stc$ of $\ncont = 1769$ contents arranged by rank on a log--log scale.  
The most popular contents reached extremely high scores, with the top five being  
1. Yo soy Betty, la fea with $62,671$ points,  
2. Bridgerton with $56,828$ points,  
3. Stranger Things with $50,364$ points,  
4. Pablo Escobar, The Drug Lord with $49,617$ points, and  
5. Squid Game with $49,527$ points.  
By contrast, a large portion of the catalog attracted very little attention, with 671 out of 1,769 titles scoring fewer than 100 points.
This highlights the highly uneven nature of global consumption,
where only a small number of titles dominate.

Panel (b) shows the distribution of $\stc$ on a log--log scale using fixed-width binning.
Each bin has a width of 10, so the first bin includes scores from 1 to 10, the second from 11 to 20, the third from 21 to 30, and so on.
The vertical axis indicates the number of contents $n(\stc)$ in each bin.
The figure clearly demonstrates a heavy-tailed pattern, where a few contents accumulate very high scores while most remain at extremely low levels.

Together, these results highlight the highly uneven distribution of global content popularity, where a small number of blockbuster shows attract overwhelming attention compared to the majority of other titles.
Hence, this analysis indicates that the top-10 lists in each country may capture enough information for the study.

\subsection{Robustness Analysis on Similarity Metrics}

\begin{figure}[!htbp]
\centering
\ff{.48}{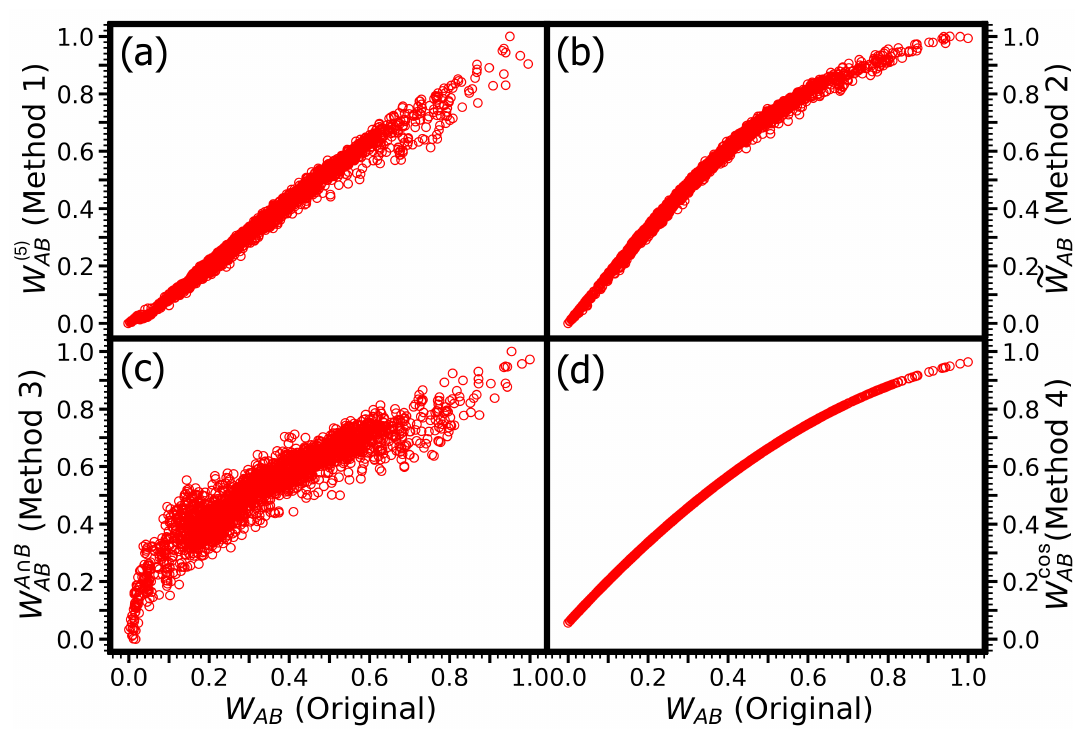}
\caption{
  Comparison of the weights $\wab$ between all $\ncoun(\ncoun-1)/2=2485$ pairs of countries
  calculated via different methods. 
Across all subfigures, the x-axis represents weights computed by the original method using Eq.~(\ref{e-wab}) in the main text.
The y-axis in each subfigure corresponds to weights recalculated using alternative methods, (a) considering only top 5 content instead of top 10 content (alt-method 1), (b) assigning a distance of 5.5 when a title was outside the top-10 in both countries (alt-method 2), (c) including only content that appeared in the top 10 list of both countries (alt-method 3), and (d) computing cosine similarity between country vectors instead of Euclidean distance (alt-method 4).
All panels show strong correlations, supporting the robustness of the original method.
}
\label{fig:figureA2}
\end{figure}

Here, we consider different ways of defining similarity weights between two countries and test whether the observed patterns remain consistent under these alternatives.  
For the four alternative methods, we present scatter plots in which the $x$-axis represents the weights computed by the original method using Eq.~(\ref{e-wab}) in the main text, and the $y$-axis corresponds to the weights recalculated using the alternative methods.  
To quantify the consistency, we calculate the Pearson correlation coefficient $R$, which measures the strength of a linear relationship between two variables and is defined as  
\beqa
 R \a= \frac{\sum_{i=1}^N (x_i - \bar{x})(y_i - \bar{y})}
            {\sqrt{\sum_{i=1}^N (x_i - \bar{x})^2} \, \sqrt{\sum_{i=1}^N (y_i - \bar{y})^2}},
\eeqa
where $x_i$ and $y_i$ are paired data points, and $\bar{x}$ and $\bar{y}$ are their respective means.  
We also report the coefficient of determination $R^2$, which measures how much of the variation in one variable can be explained by the other.  
It is defined as  
\beqa
 R^2 \a= 1 - \frac{\text{SS}_{\text{res}}}{\text{SS}_{\text{tot}}} ,
\eeqa
where $\text{SS}_{\text{res}} = \sum_{i=1}^N (y_i - \hat{y}_i)^2$ is the residual sum of squares,
and $\text{SS}_{\text{tot}} = \sum_{i=1}^N (y_i - \bar{y})^2$ is the total sum of squares.  
Here, $\hat{y}_i$ denotes the fitted value from the regression and $\bar{y}$ is the mean of the observed data.

We begin by testing whether using only the top 10 rankings is a valid approach for calculating distances between countries.
Since data beyond the top 10 is not available, all lower-ranked content is treated the same, which may overlook meaningful differences in viewer preferences.
To assess the impact of this limitation, we calculate alternative weights $\wabf$ using only the top 5 rankings (referred to as alternative method 1) and examine whether the resulting weight ordering depends on the ranking depth.
In this version, the number of content items is reduced from 1,769 to 1,314.
Scores are assigned from 5 to 1 for ranks 1st through 5th, and 0 for all other ranks.
Figure~\ref{fig:figureA2}(a) compares these similarity weights with those from the original top-10-based method.
The high correlation between the two sets of weights, with Pearson correlation coefficient $R = 0.992$ and the coefficient of determination $R^2 = 0.985$, suggests that our main results are not strongly affected by the ranking depth, supporting the robustness of our approach across different cutoff choices.

Next, we examine the implication of assigning a uniform score of 0 to content not in the top 10.
Such an assignment nullifies its contribution to distance even if the unranked content differs substantially across countries.
To approximate possible differences among unranked items, we instead assign a notional distance value of 5.5 when a title is ranked 11th or lower in both countries.
This reflects the average value of scores within the top 10, introducing a uniform middle-ground contribution to the distance measure.
Yet, the selection of 5.5 as the constant value is somewhat arbitrary but is used to illustrate that incorporating a specific, finite value for ranks beyond the top 10 does not markedly influence our findings.
Figure~\ref{fig:figureA2}(b) shows the comparison of weights under this definition (referred to as alternative method 2) and the original definition.
A strong correlation is again observed, with $R = 0.983$ and $R^2=0.967$ suggesting that the particular treatment of unranked content does not substantially alter the global similarity patterns.

The timing of content releases and the availability of specific shows vary across countries, which could influence our weight calculations.
To assess the robustness of our method to such differences, we conducted an alternative analysis using only content that simultaneously appeared in the top 10 lists of both countries on the same day.
By focusing exclusively on shared content, we eliminate potential biases caused by differences in availability.
Figure~\ref{fig:figureA2}(c) compares the resulting weights from this method (referred to as alternative method 3) with those obtained using the original definition.
Once again, we observe a quite strong positive correlation between the two sets of weights, with Pearson correlation coefficient $R = 0.926$ and $R^2=0.858$ indicating that our similarity measure remains consistent even when restricted to commonly available content.

Finally, we examine whether our results depend on the choice of distance metric itself.
All previous calculations used Euclidean distance between country vectors.
As an alternative, we compute cosine similarity between country vectors, defined as the cosine of the angle between the two high-dimensional vectors.
This similarity is often used in information retrieval and recommendation systems because it focuses on directional similarity, regardless of magnitude.
We then transform the cosine similarity values into a corresponding distance measure and calculate the weights $\wabf$ accordingly (alternative method 4).
Figure~\ref{fig:figureA2}(d) compares these weights with those from the original Euclidean-based method.
We again observe a high correlation ($R = 0.991$ and $R^2=0.982$), indicating that the overall similarity patterns among countries are not sensitive to the choice between Euclidean and cosine distance.
Together, these findings confirm the robustness of our approach under a range of plausible variations.

\subsection*{Acknowledgments}
  This work was supported by the Ministry of Education of the Republic of Korea and
  the National Research Foundation of Korea (NRF-2022S1A5A2A03051182).

\subsection*{Ethics and Consent to Participate}
Not applicable.

\subsection*{Consent to Publish}
Not applicable.

\bibliographystyle{unsrt}
\bibliography{main}

\end{document}